# Magnetoelectricity in multiferroics: a theoretical perspective

Shuai Dong[1,*], Hongjun Xiang[2,3,*] and Elbio Dagotto[4,5]

[1]*School of Physics, Southeast University, Nanjing 211189, China*

[2]*Key Laboratory of Computational Physical Sciences (Ministry of Education), State Key Laboratory of Surface Physics, and Department of Physics, Fudan University, Shanghai 200433, China*

[3]*Collaborative Innovation Center of Advanced Microstructures, Nanjing 210093, China*

[4]*Department of Physics and Astronomy, University of Tennessee, Knoxville, TN 37996, USA*
[5]*Materials Science and Technology Division, Oak Ridge National Laboratory, Oak Ridge, TN 37831, USA*

*Corresponding authors. Emails: sdong@seu.edu.cn; hxiang@fudan.edu.cn

**Abstract**

The key physical property of multiferroic materials is the existence of a coupling between magnetism and polarization, i.e. magnetoelectricity. The origin and manifestations of magnetoelectricity can be very different in the available plethora of multiferroic systems, with multiple possible mechanisms hidden behind the phenomena. In this Review, we describe the fundamental physics that causes magnetoelectricity from a theoretical viewpoint. The present review will focus on the main stream physical mechanisms in both single phase multiferroics and magnetoelectric heterostructures. The most recent tendencies addressing possible new magnetoelectric mechanisms will also be briefly outlined.

**Introduction to Magnetoelectricity & Multiferroics**

Magnetism and electricity are two fundamental physical phenomena which have been widely covered in elementary textbooks of electromagnetism and have led to a broad technological revolution within human civilization. Even today, these two crucial subjects remain at the frontier of active research, and are still attracting considerable attention within the scientific community for their indispensable scientific value and possible applications. In solids, magnetism and electricity originate from the spin and the charge degrees of freedom, respectively. The crossover between these two fascinating topics has much grown in recent years and it has developed into an emergent branch of Condensed Matter Physics called *magnetoelectricity* [1-9].

Generally speaking, magnetoelectric effects can exist in many systems, even in some that are nonmagnetic. In fact, the first example of a magnetoelectric effect was observed by Röntgen in 1888 in a dielectric material, which was magnetized when moving through an electric field [10]. Much more recently, the surface state of topological insulators was predicted to manifest magnetoelectric effects [11]. However, to develop magnetoelectricity of a large magnitude, and as a consequence of more considerable practical value, multiferroics seem to be the best playground. In multiferroics, both magnetic moments and electric dipole moments can be ordered, inducing robust macroscopic quantities such as magnetization and polarization. Moreover, crucially for applications, *both moments are coupled.* Then, these macroscopic quantities may be mutually controlled, for example modifying the magnetization by an electric voltage or modifying the polarization by a magnetic field, which is particularly useful to design new devices, such as for storage and sensors.

However, conceptually the mere existence of multiferroics is highly non trivial [12]. For most magnetic materials, the magnetic moments arise from unpaired electrons in partially occupied $d$ orbitals and/or $f$ orbitals. However, the spontaneous formation of a charge dipole usually needs empty $d$ orbitals as a condition of having a coordinate bond, i.e. the so-called $d^0$ rule. Thus, the key ions involved in typical magnetic materials and those in polar materials are different, making these two areas of research nearly isolated from each other. However, in 2003 the discovery of a large polarization in a $BiFeO_3$ film [13] and magnetism-induced polarization in a $TbMnO_3$ crystal [14] opened the new era of multiferroic materials. Accompanying the subsequent rapid bloom of the multiferroic field, the theories of magnetoelectricity developed fast as well, and have become more and more complete.

As stated before, due to their different origins, it is nontrivial to couple magnetism and electric polarity together in solids. In spite of this conceptual complication, research in the past few years have found several "glues" that may link these apparently disjoint phenomena.

The first "glue" is provided by the spin-orbit coupling, a relativistic effect. In principle, a charge dipole breaks the space inversion symmetry, while a spin breaks the time reversal symmetry. Time and space are independent in non-relativistic physics, but the relativistic effect can link time and space. Thus, the spin-orbit coupling may link magnetic moments and charge dipoles. In particular, some nontrivial magnetic textures, such as magnetic orders with chirality, can break the space inversion symmetry. Then, the spin-orbit coupling can translate this symmetry breaking into a charge dipole, as it occurs in TbMnO$_3$ [15]. Conversely, if the space inversion is broken, the spin-orbit coupling can control the texture of magnetic moments, as it occurs in BiFeO$_3$ [16]. Usually, noncollinear spin textures are always associated with magnetoelectricity mediated by the spin-orbit coupling.

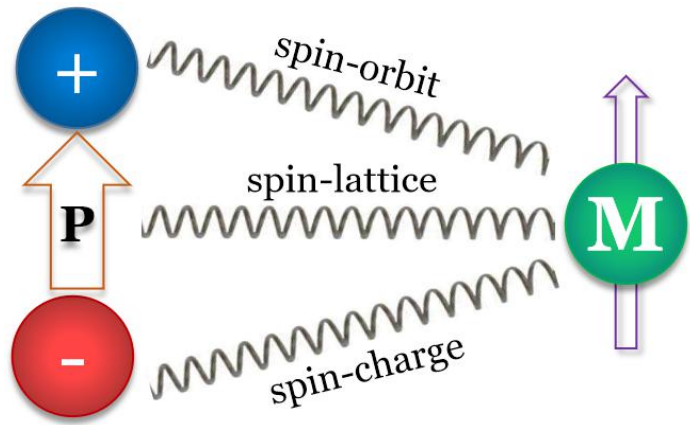

**Fig. 1**. Schematic drawing of possible magnetoelectric couplings. Left: a charge dipole indicative of ferroelectricity. Right: a magnetic moment indicative of magnetism. Three "glues" are shown that can link these two vectors.

The second "glue" is the spin-lattice coupling. The magnetic interactions between magnetic ions, both the regular symmetric exchanges and the antisymmetric Dzyaloshinskii-Moriya interaction, depend on the details of the electronic exchange paths. Microscopically, the changes of bond angles and lengths seriously affect the overlaps between wave functions and, thus, the exchanges. Macroscopically, the expression magnetostriction effects refer to the changes of the sample's shape under magnetic fields or upon magnetic ordering, an effect that has been known for many years for magnetic materials. Furthermore, the single-site magnetocrystalline anisotropy also depends on the crystalline field, which can be tuned by the lattice distortions. For multiferroics with both magnetism and polarity, such magnetostriction effects establish a link between the polarization and the magnetism.

The third "glue" is the spin-charge coupling, mediated by the charge density distribution [17]. Since carriers (electrons or holes) can be spin-polarized in magnetic systems, the local magnetization (or even the magnetic phases) can be tuned by modulating the charge density distribution [18]. Both external electric fields, ferroelectric fields, and polar interfaces [19] can be the driving force that moves the carriers.

Although the aforementioned three "glues" are classified based on three different degrees of freedom of electrons in solids, in many cases these three "glues" cooperate together. One may act as the primary driving force, while the others play a secondary role. Thus, to fully understand the magnetoelectricity in multiferroics, it is necessary to carefully analyze the possible underlying mechanisms. In the following, we will briefly introduce some concrete examples to illustrate these dominant magnetoelectric couplings.

## Magnetoelectricity in concrete multiferroic systems

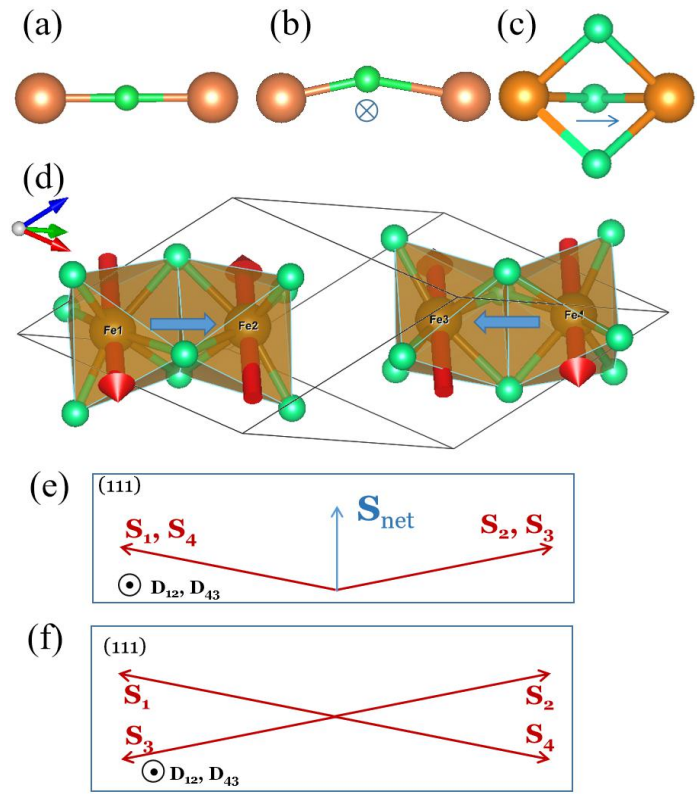

**Fig. 2.** Schematic representation of the Dzyaloshinskii-Moriya interaction. Brown spheres: magnetic ions; Green spheres: anions. (a) The anion is an inversion center. Thus the **D** vector is zero. (b) There are two mirror planes, one bisecting and one passing through the line connecting two magnetic ions. The **D** vector is perpendicular to the plane of these three ions. (c) The line connecting two magnetic ions is a triple-fold axis, thus the **D** vector is along this line. (d) Crystalline structure of $\alpha$-$Fe_2O_3$. The room-temperature spin order of irons is shown as red arrows. Blue arrows: $\mathbf{D}_{12}$ and $\mathbf{D}_{34}$. (e) The spin canting due to the Dzyaloshinskii-Moriya interaction in $\alpha$-$Fe_2O_3$. The spins are in the (111) plane while the **D** vector is along the [111] axis. A net magnetization will be induced, as indicated by $\mathbf{S}_{net}$. (f) For the isostructural $Cr_2O_3$ with different magnetic order, the canting moment is canceled.

*A. Role of Dzyaloshinskii-Moriya interaction*

The Dzyaloshinskii-Moriya interaction frequently plays a vital role for magnetoelectricity in various multiferroics. It was first proposed by Dzyaloshinskii in 1958 to explain phenomenologically the weak ferromagnetism observed in α-$Fe_2O_3$ [20]. Driven by the Dzyaloshinskii-Moriya interactions, the antiferromagnetically ordered spins in α-$Fe_2O_3$ become canted by a small amount, leading to a residual net magnetization. Later, Moriya further clarified its origin at the microscopic level [21, 22]. Despite its complex origin from spin-orbit coupling, its final form can be elegantly expressed as:

$$H_{DM}=\mathbf{D}_{ij}\cdot(\mathbf{S}_i\times\mathbf{S}_j), \quad (1)$$

where **S** represents a spin vector and **D** is a coefficient vector. According to this expression, it is natural to expect that (i) spin pairs become noncollinear due to the cross product between spins and (ii) there is a spin-orientation dependence due to the **D** vector which is fixed. Then, it is crucial to know the orientation of **D**, which depends on the crystalline symmetries. Based on symmetry analysis, Moriya figured out five helpful rules to determine the orientation of $\mathbf{D}_{ij}$ between the spins located at sites *i* and *j* [21]:

1) If the bisecting point of *i* and *j* is an inversion center, then **D**=0 (Fig. 2(a)).

2) If there is a mirror plane perpendicular to the line *i-j*, **D** is also perpendicular to the line *i-j* (Fig. 2(b)).

3) If there is a mirror plane passing through *i* and *j*, **D** is perpendicular to this mirror plane (Fig. 2(b)).

4) If there is a two-fold rotation axis perpendicular to the line *i-j*, then **D** is perpendicular to this axis.

5) If the line *i-j* is an *n*-fold axis (*n*≥2), **D** is along the line *i-j* (Fig. 2(c)).

Considering α-$Fe_2O_3$ as example, see Fig. 2(d), the line Fe1-Fe2 (and Fe3-Fe4) is a triple-fold axis, thus $\mathbf{D}_{12}$ (and $\mathbf{D}_{34}$) is along the [111] axis. However, for Fe2-Fe3 (and Fe1-Fe4), the bisecting point is the inversion center, thus both $\mathbf{D}_{23}$ and $\mathbf{D}_{14}$ are zero. The inversion center between Fe2-Fe3 also requires that the sign of $\mathbf{D}_{12}$ and $\mathbf{D}_{34}$ must be opposite. In summary, for α-$Fe_2O_3$, $\mathbf{D}_{12}=-\mathbf{D}_{21}=-\mathbf{D}_{34}=\mathbf{D}_{43}$, and $\mathbf{D}_{23}=-\mathbf{D}_{32}=-\mathbf{D}_{41}=\mathbf{D}_{14}=0$. The spins of irons $S_{1-4}$ are almost +--+, e.g. $S_1\sim-S_2\sim-S_3\sim S_4$, pointing perpendicular to the [111] axis (at room temperature). Then, the Dzyaloshinskii-Moriya interaction can drive the spin canting between $S_1$ and $S_2$ (or $S_3$ and $S_4$). As shown in Fig. 2(e), the canting directions are identical for $S_1$-$S_2$ and $S_3$-$S_4$, leading to a net magnetization in the (111) plane, i.e. a weak ferromagnetism.

At low temperature (<250 K), the spins in α-$Fe_2O_3$ reorient to the [111] axis, parallel or antiparallel to the **D** vectors. Then, the Dzyaloshinskii-Moriya interaction cannot lead to spin canting anymore and, thus, the weak ferromagnetism disappears [19]. The case of $Cr_2O_3$ is a little different: while its crystalline structure is identical to α-$Fe_2O_3$ its spin order is instead +-+- for $S_{1-4}$. Although these spins are lying in the (111) plane, the canting effect driven by the Dzyaloshinskii-Moriya interaction cannot lead to a net magnetization [20], as shown in Fig. 2(f). However, the asymmetric configuration ($D_{12}$=-$D_{34}$) can be broken by applying an electric field (**E**), which can slightly distort the structure and, thus, break the symmetry. Then, a net magnetization (**M**) emerges, as a linear magnetoelectric response, i.e. **M**~$α_{ME}$**E**, where $α_{ME}$ is the magnetoelectric coefficient.

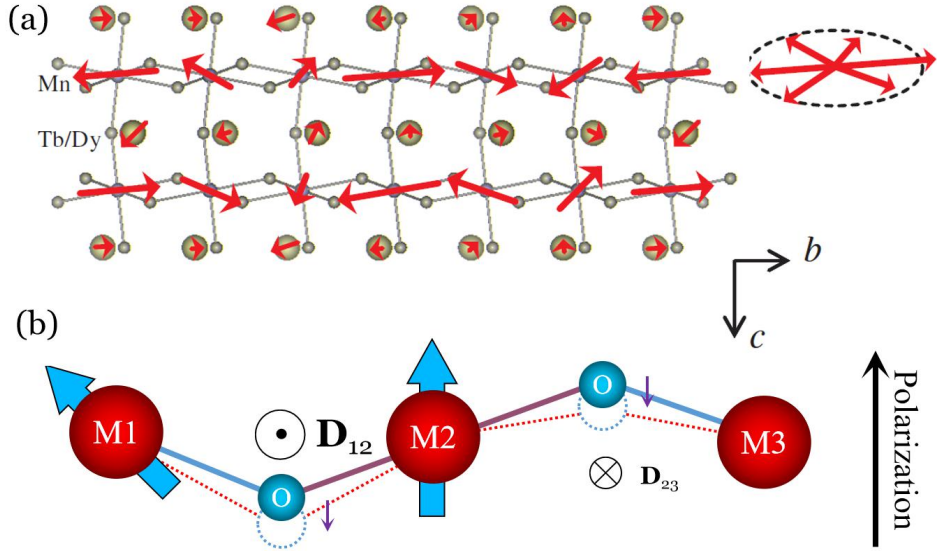

**Fig. 3**. Schematic drawing of magnetism-driven electric polarization in TbMnO$_3$. (a) The spins (magnetic moments) of Mn form a (distorted) cycloid order, which lies in the *bc* plane and propagates along the *b*-axis. Right: the trajectories of the Mn spins along the cycloid. Reprinted figure with permission from Arima *et al*. [23]. Copyright©(2006) by the American Physical Society. (b) Due to the GdFeO$_3$-type distortion, the original Dzyaloshinskii-Moriya vectors are staggered between nearest neighbors, i.e. **D**$_{12}$=-**D**$_{23}$. However, the cycloid magnetic order prefers the same orientation of the Dzyaloshinskii-Moriya vectors to reduce the energy, which is superimposed on the original staggered pattern. Since the magnitude of the Dzyaloshinskii-Moriya vector is proportional to the Mn-O-Mn bond bending, the bias of the Dzyaloshinskii-Moriya vectors leads to unidirectional displacements (indicated by small purple arrows) of the oxygen ions, leading to a net polarization. By changing the helicity and plane of the cycloid spins, the polarization can be modulated accordingly.

Indeed, $Cr_2O_3$ was the first experimentally confirmed magnetoelectric material, although it is not a multiferroic material. However, the "glue" role of the Dzyaloshinskii-Moriya interaction is similar. Considering the so-called type-II multiferroics, such as TbMnO$_3$ for example, the Dzyaloshinskii-Moriya interaction is the most important ingredient for its magnetoelectricity [24]. As shown in Fig. 3(a), the spins of Mn form a cycloid order, lying in the ***b-c*** plane and propagating along the *b*-axis [22, 25, 26]. This cycloid spin texture, with unidirectional ($S_i \times S_j$)||$a$, will drive a bias of $D_{ij}$ along the $a$-axis. According to Moriya's rules, the bending of each Mn$_i$-O-Mn$_j$ bond breaks the inversion center and leads to a finite $D_{ij}$ perpendicular to the Mn$_i$-O-Mn$_j$ plane [22], as sketched in Fig. 2(b). The reverse effect is that a biased $D_{ij}$ drives a biased bending of the Mn$_i$-O-Mn$_j$ bonds. At the first order approximation of a Taylor expansion, the value of $D_{ij}$ is linearly proportional to the movement of the O ion from the bond center (i.e. the original inversion point) [24]. It should be noted that the original Mn-O-Mn bonds are already seriously bended due to the collaborative tilting and rotation of the oxygen octahedra (the so-called GdFeO$_3$-type distortion for the perovskite structure of *Pbnm* group), independently of the magnetic properties. Thus, the biasing of the vectors $D_{ij}$'s due to the cycloid order leads to additional unidirectional displacements of the O ions in the *b*-*c* plane (Fig. 3(b)). Considering the propagation direction of the cycloid order to be along the *b*-axis, the net induced polarization is along the *c*-axis. In summary, the (inverse effect of) the Dzyaloshinskii-Moriya interaction is the engine used by the noncollinear magnetism to generate a net electric polarization [24]. Those multiferroics with such physical process were vividly coined as "quantum electromagnets" by Tokura [27].

Such inverse effect of the Dzyaloshinskii-Moriya interaction can also be interpreted using the spin-current model, i.e. the so-called Katsura-Nagaosa-Balatsky (KNB) model [28], which leads to a similar expression:

$P \sim e_{ij} \times (S_i \times S_j)$, (2)

where $P$ is the polarization and $e_{ij}$ is the unit vector linking the two spins. The derivation of this model is based on the use of perturbation theory applied to the Hubbard model with spin-orbit coupling. A conclusion similar to Eq. 2 can also be derived via a phenomenological model based on the Landau free energy [29]. Readers are referred to the original publications for more details.

TbMnO$_3$ is a representative of type-II multiferroics, whose mechanism of magnetoelectricity can be applied to many other type-II multiferroics with noncollinear spin orders. More details of related materials and their physical properties can be found in the following reviews [7, 15, 30].

By contrast, BiFeO$_3$ is a typical representative of type-I multiferroics, whose polarization does not originate from a nontrivial magnetic texture. As the most studied multiferroic material, BiFeO$_3$ displays prominent properties, including a large polarization (~90 μC/cm$^2$ in the rhombohedral phase [13, 31] and even a larger one in the tetragonal phase [32]) as well as ferroelectric and magnetic ordering above-room temperatures. Its large polarization arises from the $6s^2$ lone pair of Bi$^{3+}$ ions [33], avoiding the $d^0$ rule restriction for magnetism. Moreover, the magnetic ordering of the Fe$^{3+}$ ions can also be robust. In this sense, BiFeO$_3$ can be considered as an atomic-level magnetoelectric "composite". However, the magnetoelectricity of BiFeO$_3$ remains dominated by the Dzyaloshinskii-Moriya interaction [16]. The canting moment of each antiferromagnetic spin pair is the direct result of the Dzyaloshinskii-Moriya interaction [16], like the physics in the aforementioned α-Fe$_2$O$_3$ case. This canting moment forms a long-periodic cycloid modulation, canceling the net magnetization [34]. In spite of this, considering a pair of two Fe sites for example, a simplified equation for such magnetoelectricity can be expressed as: **M**~**P**×**L**, where **M** is the local magnetic moment generated by spin canting, **P** is the polarization, and **L** is the antiferromagnetic order parameter (defined as **S**$_1$-**S**$_2$ where **S** is a spin) [16]. Such magnetoelectricity leads to a perpendicular relationship between the polarization and the magnetic easy plane, as well as a correspondence between ferroelectric and antiferromagnetic (weak ferromagnetic) domains [35, 36]. In addition, the spin-charge coupling may also contribute to the magnetoelectricity present in BiFeO$_3$ domain walls and interfaces with others materials, since the head-to-head/tail-to-tail domain wall and interfaces are polar discontinuous, which can trap carriers [37]. Such a mechanism will be explained in more detail later. Experimentally, multiple magnetoelectric couplings have been demonstrated, most of which are ferroelectric domain related. The microscopic physical mechanism can be complicated and the net result may arise from more than one mechanism. More details of the magnetoelectricity in BiFeO$_3$ can be found in the following reviews [36, 38].

Besides the materials TbMnO$_3$ and BiFeO$_3$, the hexagonal manganites $R$MnO$_3$ (also the hexagonal ferrites $R$FeO$_3$) and the 327-series Ruddlesden-Popper perovskites are much studied multiferroics in recent years. They are improper ferroelectrics, but their ferroelectricity arises from the cooperation of multiple structural distortional modes [39]. For example, in hexagonal $R$MnO$_3$ (or in $R$FeO$_3$), the tilting of the oxygen bipyramids and the trimerization of the Mn (Fe) triangles generate the uncompensated displacements of the $R$ ions along the $c$-axis [40], as shown in Fig. 4(a). The ferroelectric Curie temperatures are considerably high (much higher than room temperature in most members) and the polarization remains moderate (typically ~10 μC/cm$^2$). The magnetic moments of Mn (Fe) become ordered usually at low temperatures (~100 K for Mn and a little higher for Fe) [41, 42]. The magnetic moments of Mn (or Fe) lie in the *a-b* plane, forming the non-collinear Y-type antiferromagnetism due to the exchange frustration of the triangular lattice geometry [43], as shown in Figs. 4(b-c). Then, the bulk magnetoelectricity can be obtained with the help of the Dzyaloshinskii-Moriya interaction. The ferroelectric polar structure, i.e. the trimer distortion, induces a transverse component (in the *a-b* plane) of the Dzyaloshinskii-Moriya vector, which leads to a tiny canting of magnetic moments along the *c*-axis, i.e. a net magnetization [43]. In principle, using an electric field applied along the *c*-axis to modulate the polarization (e.g. buckling of MnO$_5$ polyhedra), the transverse component of the Dzyaloshinskii-Moriya interaction can be tuned. Thus, a magnetoelectric response can be expected.

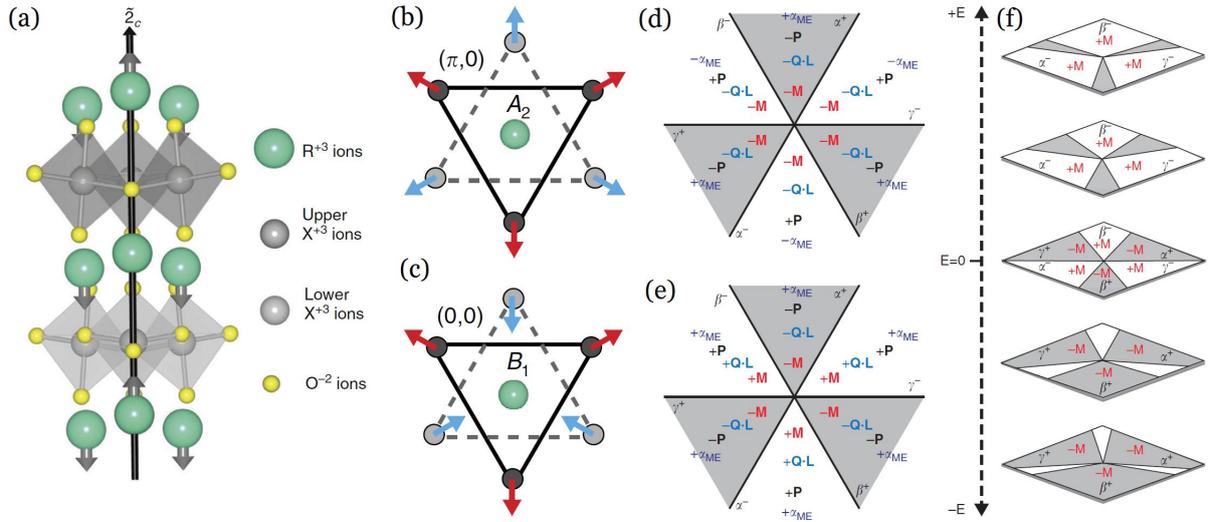

**Fig. 4.** Magnetoelectricity in hexagonal $R$MnO$_3$ and $R$FeO$_3$. (a) Schematic of the crystal structures. The displacements of the $R^{3+}$ ions are indicated by arrows, leading to a net polarization. (b-c) The in-plane (*ab* plane) geometry of the Mn (or Fe) sites. The solid and broken triangles denote the upper and lower layers within one unit cell. The Y-type 120º noncollinear antiferromagnetic textures are shown. According to density functional theory calculations, the ground state is (b) $A_2$ for LuFeO$_3$, but (c) $B_1$ for LuMnO$_3$. The Dzyaloshinskii-Moriya interaction can induce a slight canting of these magnetic moments to the +*c*/-*c* axis. Only the $A_2$ configuration can present a net magnetization, while the magnetization is canceled between layers for the $B_1$ case. In spite of these distinctions, the energy difference between $A_2$ and $B_1$ is very small (due to the spin-orbit coupling), which can be overcome by external magnetic fields. (d-e) Schematic of six-fold topological ferroelectric/structural antiphase/antiferromagnetic domains for the $A_2$ phase. *Q*: phase of the structural trimer distortion; **L**: antiferromagnetic order; **P**: polarization; **M**: net magnetization. Across the domain wall, the antiferromagnetic spins rotate by (d) ±π/3, or (e) ±2π/3. In case (d), the induced canting magnetization does not change between ferroelectric domains, but the sign of the magnetoelectric coefficient ($\alpha_{ME}$) changes. In case (e), the net magnetization clamps to the ferroelectric domain, but the sign of the magnetoelectric coefficient ($\alpha_{ME}$) does not change. In this case, the electric field can not only tune the ferroelectric domains but also the net magnetization, as sketched in (f). Reproduced with permission from Macmillan Publishers Ltd: Das *et al*. Ref. [43]. Copyright©(2014).

The most interesting issue in these hexagonal systems is the domain-related magnetoelectricity. The polarization of hexagonal manganites/ferrites is bi-valued (up or down), while the antiphase structural domains due to the trimerization are triple-valued (*α*, *β*, *γ*) [44]. Due to the complex coupling between the ferroelectric distortion and structural trimerization [45, 46], special topological domain structures with $Z_2 \times Z_3$ vortices/anti-vortices are formed [44, 47], as sketched in Figs. 4(d-f). Across domain walls, the antiferromagnetic spins rotate by ±π/3 or ±2π/3 [43]. In the ±π/3 case, the induced canting magnetization does not change between the two ferroelectric domains, but the sign of the magnetoelectric coefficient ($\alpha_{ME}$) changes. Experimentally, by applying a high magnetic field to align the canting magnetization in all domains, the sign of the magnetoelectric coefficient indeed changes with the ferroelectric domains [48], implying magnetoelectric domains, as shown in Fig. 4(d). In the $\pm 2\pi/3$ case, the sign of $\alpha_{ME}$ is fixed, then the direction of the local magnetization follows the sign of the polarization, as shown in Fig. 4(e), which remains to be verified experimentally. Although the domain vortex cannot be easily erased, the ferroelectric domain size can be tuned by an electric field, as shown in Fig. 4(f), which will affect the value of $\alpha_{ME}$ or the local magnetization.

In summary, the Dzyaloshinskii-Moriya interaction, stemmed from the spin-orbit coupling and associated with noncollinear spin textures, plays a vital role for magnetoelectricity in many multiferroics, not only in type-II multiferroics but also in many cases of type-I multiferroics, as well as in heterostructures [49, 50].

*B. Role of symmetric exchange*

Although the aforementioned Dzyaloshinskii-Moriya interaction originates in the relativistic spin-orbit coupling, lattice distortions often occurs in most situations, which determine the directions and magnitudes of the Dzyaloshinskii-Moriya vectors. Thus, more strictly, the aforementioned magnetoelectricity is based on spin-orbit-lattice coupling. In this sub-section, the pure spin-lattice coupling without the relativistic effect in multiferroics will be described, which is called the symmetric exchange striction.

The best example to illustrate the symmetric exchange striction is $Ca_3CoMnO_6$, in which the Co and Mn ions form quasi-one-dimensional chains arranged as ...-Co-Mn-Co-Mn-... [51], as shown in Figs. 5(a-b). The magnetic moments form up-up-down-down patterns within the chain. Thus, the symmetry between the Co(up)-Mn(up) pair and the Mn(up)-Co(down) pair is broken. To gain more exchange energies, the distance between Co and Mn with parallel spins shrinks, while the distance between Co and Mn with antiparallel spins increases [52, 53]. Such displacements of ions generate a polarization along the chain direction, i.e. along the *c*-axis. Such magnetism-driven polarization does not rely on the weak spin-orbit coupling, but on the exchange interactions which can be much stronger. So in principle, the polarization generated in this manner can be (usually one order of magnitude) larger than those generated by the spin-orbit coupling [52, 53]. However, since the polarization is only related to the inner product $\mathbf{S}_i \cdot \mathbf{S}_j$, the particular directions of magnetic moments are not involved. Thus, the magnetoelectric response is typically weak in multiferroics with this mechanism. Moreover, a large enough magnetic field can suppress the -up-up-down-down- type antiferromagnetism, leading to a -up-up-up-down- ferrimagnetic or even full ferromagnetic states, in which the polarization should also be suppressed [54].

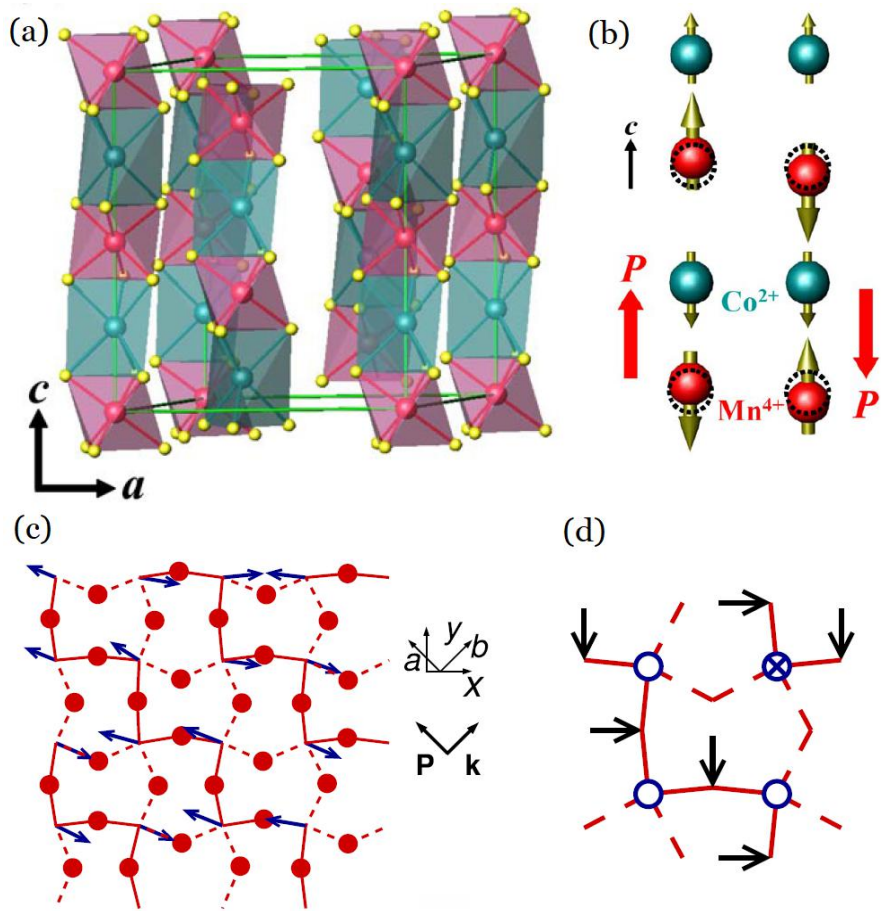

**Fig. 5.** Schematic of symmetric exchange striction. (a) Crystalline arrangement of $Ca_3CoMnO_6$ with quasi-one-dimensional ...-Mn-Co-... chains along the *c*-axis. (b) With the up-up-down-down magnetic order, the Co-Mn distances are distorted. For reference, the original positions before exchange striction are shown as dashed circles. These distortions lead to a net polarization along the *c*-axis, which can be switched by changing the phase of the antiferromagnetic order. (a-b) Reprinted figure with permission from Choi *et al*. [51]. Copyright©(2008) by the American Physical Society. (c) In-plane crystalline structure of perovskite $HoMnO_3$. Arrows: Mn's spins; red circles: oxygen ions. The solid lines connect parallel spins while the broken lines connect antiparallel spin pairs. The Mn-O-Mn bonds are straighter for the parallel spin pairs along the zigzag chain, as emphasized in (d). In (d) Mn sites are shown as circles. (c-d) Reprinted figure with permission from Sergienko *et al*. [55]. Copyright © (2006) by the American Physical Society.

Such exchange striction mediated magnetoelectricity works in many multiferroics, such as orthorhombic $HoMnO_3$ with E-type antiferromagnetism [55, 56] [see Figs. 5(c-d)] and iron selenides $BaFe_2Se_3$ [57]. Even in some prototypes of cycloid magnets, like $DyMnO_3$, $Eu_{1-x}Y_xMnO_3$, as well as $CaMn_7O_{12}$, the symmetric exchange strictions also take part in between Mn-Mn or Dy-Mn, which can enhance the net polarization [58-63].

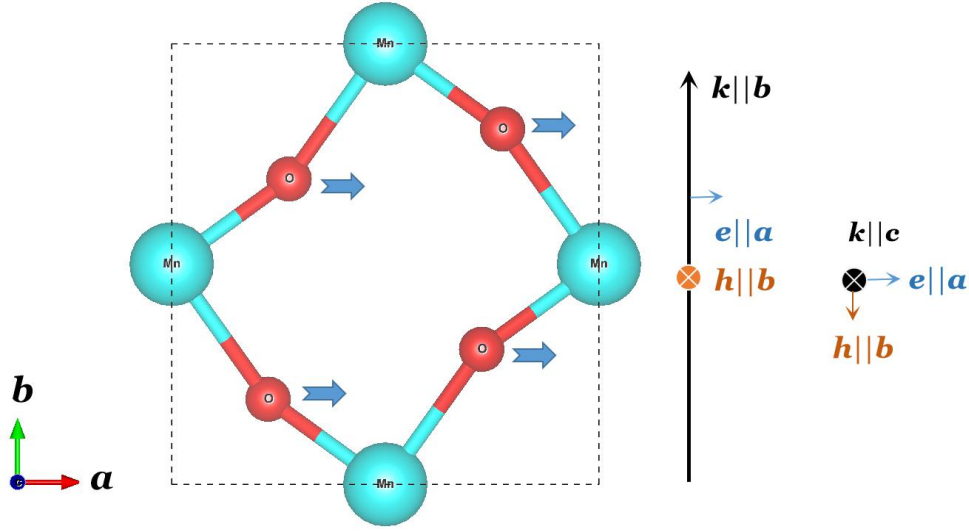

**Fig. 6.** Schematic mechanism for the generation of electromagnons in perovskite manganites. The *ab*-plane crystalline structure is shown, with bending Mn-O-Mn bonds in the GdFeO$_3$-type manner. Arrows denote the instantaneous displacements of the oxygen ions driven by the electric field component of light, which dynamically make half of the Mn-O-Mn bonds more bending and the other half more straight. Such dynamic modulation disturbs the magnetic ground state, i.e. it generate magnons. This process is orientation selective. Only with a nonzero electric field component (*e*) along the *a*-axis (right panel), the electromagnons can be generated. *h* and *k* denote the magnetic field component and wave vector of light, respectively. The typical photon energy to excite electromagnons in perovskite manganites (such as Eu$_{1-x}$Y$_x$MnO$_3$) is in the THz range [64, 65].

Besides the aforementioned static magnetoelectricity, dynamic magnetoelectricity can also be driven by symmetric exchange striction. An important conceptual issue to address for dynamic magnetoelectricity is the so-called electromagnon, i.e. the possibility of exciting magnons using *a.c.* electric fields. Experimentally, THz electromagnetic waves can be absorbed by GdMnO$_3$, TbMnO$_3$, and Eu$_{1-x}$Y$_x$MnO$_3$ with spiral spin orders [64, 65], with a selected direction of the electric field component, e.g. **E**∥*a*. The mediator is the vibration of the Mn-O-Mn bond distortions, which is driven by the electric field component and then modulates the magnetic exchanges, as shown in Fig. 6. In addition to the cycloidal phase, a colossal electromagnon excitation has also been observed in the E-type antiferromagnetic phase of TbMnO$_3$ under pressure [66]. Finally, it should be mentioned that the spin-orbit coupling can also contribute to the electromagnons but typically at a weaker level and with a different selection rule [67, 68].

*C. Role of charge modulation*

The electronic carrier density is among the most important parameters to determine the physical properties of solids. There are several routes for charge to tune the magnetoelectric properties.

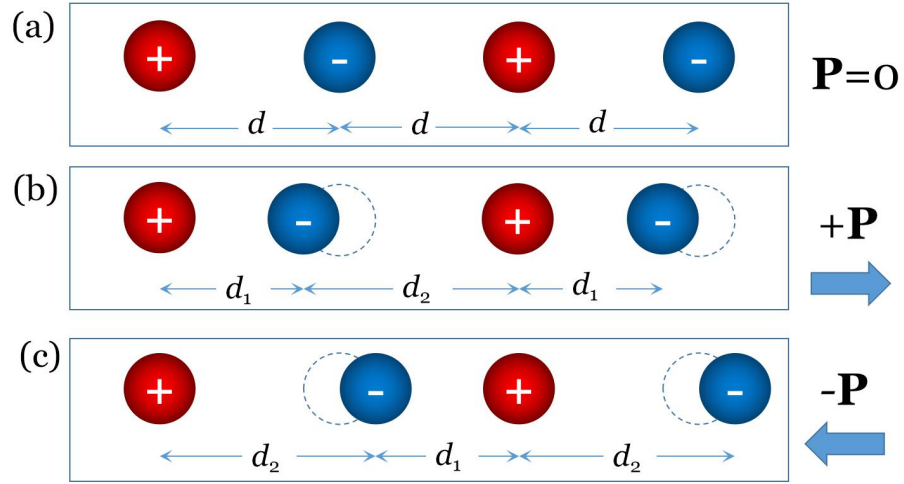

**Fig. 7.** Schematic of charge-ordering induced polarization in a one-dimensional chain. (a) An ionic chain with equivalent distance between neighbor sites. Each site is an inversion center. Thus, the chain, with periodic boundary conditions for simplicity, is non-polar. (b-c) Ionic chains with distortions. The distances between neighboring sites is nonequivalent now. Thus, there is no inversion center anymore. The polarization is $Q·(d_2-d_1)/[2·(d_1+d_2)]$, where $Q$ is the charge of the cation. The direction of polarization can be switched between (b) and (c).

First, there is one branch of ferroelectrics called electronic ferroelectrics [69]. Many transition metal ions have multiple valences, and sometimes mixed valences of a single element coexist in the same material. If the mixed-valent ions form a charge ordering pattern which breaks the space-inversion symmetry, an improper ferroelectric polarization is induced, as illustrated in Fig. 7. Usually, structural dimerization is essential for these electronic ferroelectrics [70]. Magnetism usually exists in these systems due to the contribution of the transition metals in the chemical formulas. In principle, there is no explicit relationship between the magnetism and polarization. However, usually both of them depend on the charge ordering. For example, in a recent theoretical work, trirutile $LiFe_2F_6$ is predicted to be a multiferroic, whose polarization is due to the charge ordering while its ferrimagnetic magnetization is also due to charge ordering [71]. Then, the switching of polarization provides the opportunity to synchronously flip the magnetization, leading to a strong magnetoelectric coupling. Most multiferroics are antiferromagnetic, which are not ideal for applications due to the absence of a net magnetization. In contrast, these electronic ferroelectrics are sometimes ferrimagnetic due to the uncompensated magnetic moments of charge ordering. However, the common weakness for multiferroics in this category is having too small band gaps, which lead to serious leakage preventing the ferroelectric measurements [72]. It seems that the charge ordering arising from Mottness cannot open a big gap.

Second, the ferroelectric polarization can act as an electric field at interfaces or domain walls, namely the "field effect", as sketched in Figs. 8(a-b). Then, this field effect itself can tune the local charge density [74], as it occurs in semiconductor transistors. If a magnetic material is involved, the local magnetization may be tuned as a function of the local charge density [75, 76]. In principle, the intensity of the field effect is proportional to the change of polarization, i.e. to $\nabla \cdot \mathbf{P}$, whose dimensional units are just the charge. For example, for a typical ferroelectric perovskite, the perpendicular polarization of 10 μC/cm$^2$ is equivalent to the area density of ~0.1 electron per u.c., if the lattice constant of the u.c. is ~4 Å. In other words, 0.1 extra electron (or hole) per u.c. area can fully screen the field effect of a perpendicular polarization of 10 μC/cm$^2$. The distribution length of the extra carriers, i.e. the screening length, depends on the local carrier density, which can be long (e.g. several nanometers) in semiconductors but should be very short (~ 1 u.c.) in metallic materials [18]. In an ideal limit, the modulation of the local magnetization equals to the extra carrier since one electron carries one Bohr magneton.

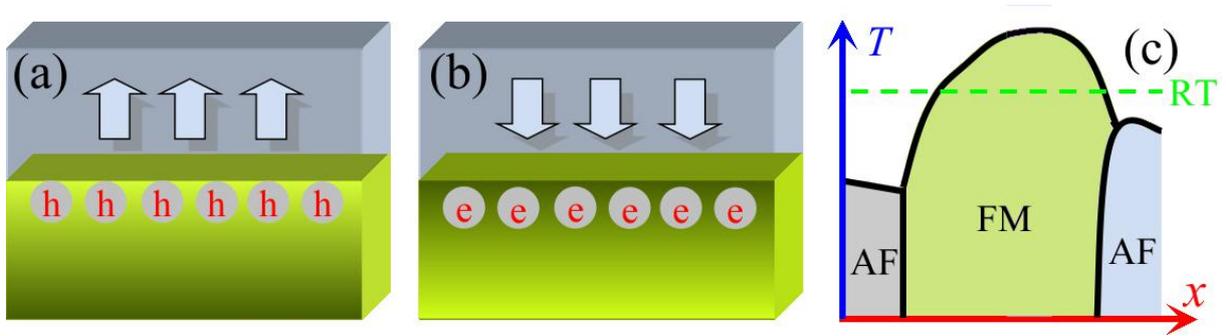

**Fig. 8.** (a-b) Schematic of ferroelectric field effect at an interface. h: hole; e: electron; arrows: direction of polarization. Once the lower layer is magnetic, the interfacial magnetization can be tuned. (c) Schematic of a magnetic phase diagram with competing phases. The magnetic phases depend on both temperature (*T*) and charge density (*x*). Thus, the ferroelectric field effect may tune the magnetic ground state, leading to a large modulation of magnetization. RT is room temperature. Reprinted figure with permission from Dong *et al*. [73]. Copyright©(2013) by the American Physical Society.

Third, this field-effect driven magnetoelectricity can be magnified if a magnetic phase transition is involved [77]. Colossal magnetoresistive manganites, which display a plethora of magnetic phases in their phase diagram (see Fig. 8(c) for example) [78], are often used to demonstrate this effect [79]. By fine tuning of doping concentration, the ground state of a few layers of a manganite can be close to a certain magnetic phase boundary. Then, the modulation of the local charge density may trigger a phase transition between antiferromagnetic (or ferrimagnetic) and ferromagnetic orders [73, 80]. For this reason, the magnitude of the change in the magnetization can be larger than 1 $\mu_B$/electron.

In summary, the modulation of the charge density in multiferroics and magnetoelectric heterostructures can lead to strong magnetoelectric effects.

*D. Other magnetoelectric mechanisms in multiferroics*

Although in most cases the magnetoelectric behaviors in multiferroics can be attributed to the three mechanisms described above, or their combinations, there are some exceptions.

The first example are some type-II multiferroics with layered triangular lattices. Two dimensional triangular lattices are typically geometrically frustrated lattices for antiferromagnets, such as $CuFeO_2$, $Ba_3NiNb_2O_9$, and a number of isostructures [81-84]. Usually the 120º non-collinear spin order (Y-type antiferromagnetism) is stabilized by the nearest-neighbor antiferromagnetic exchanges, if the second nearest-neighbor exchange and the magnetocrystalline anisotropy are weak [85]. Such non-collinear spin texture can induce a tiny polarization perpendicular to the spin plane [81-84]. The microscopic driving force is the spin-orbit coupling but not the Dzyaloshinskii-Moriya interaction. For the spin-current model, it is straightforward to obtain a zero polarization for each triangle involving sites 1, 2, 3: $\mathbf{e}_{12} \times (\mathbf{S}_1 \times \mathbf{S}_2) + \mathbf{e}_{23} \times (\mathbf{S}_2 \times \mathbf{S}_3) + \mathbf{e}_{31} \times (\mathbf{S}_3 \times \mathbf{S}_1) = 0$.

Xiang *et al.* proposed a unified polarization model to explain this magnetism driven polarization [86-90]. According to this model, the total polarization can be written as a sum of two contributions:

$$\mathbf{P} = \mathbf{P}_e(\mathbf{S}; \mathbf{U}=0, \eta=0) + \mathbf{P}_{lat}(\mathbf{U}, \eta) \quad (3)$$

where $\mathbf{P}_e$ is the electronic contribution induced by a spin order $\mathbf{S}$, and $\mathbf{P}_{lat}$ is the lattice contribution due to the ionic displacements $\mathbf{U}$ and strain $\eta$ induced by the spin order $\mathbf{S}$. Due to time-reversal symmetry, $\mathbf{P}_e = \sum_{i,\alpha\beta} \mathbf{P}^i_{\alpha\beta} S_{i\alpha} S_{i\beta} + \sum_{<i,j>,\alpha\beta} \mathbf{P}^{ij}_{\alpha\beta} S_{i\alpha} S_{j\beta}$, where the first term is the single-site term, while the second term is the intersite term. The intersite term includes a general spin current contribution [86], a symmetric exchange striction contribution, and an anisotropic symmetric exchange contribution [89]. To obtain $\mathbf{P}_{lat}$, the total energy $E(\mathbf{S}; \mathbf{U}, \eta)$ should be minimized with respect to the ionic displacements $\mathbf{U}$ and strain $\eta$ for a giving fixed spin order $\mathbf{S}$ [88]. The interacting parameters needed in the unified model can be computed with the four-states method [91].

Not only in these Y-type antiferromagnetic systems is that the spin-orbit coupling can contribute more to the magnetism-driven polarization than merely via the Dzyaloshinskii-Moriya interaction. In fact, at least there are two more microscopic mechanisms that have been identified. One is the spin-dependent metal-ligand *p-d* hybridization (a special single-site term of the unified polarization model [86]), which can be written in a formula as [92]:

$$\mathbf{P} = \sum_i (\mathbf{e}_i \cdot \mathbf{S}_i)^2 \mathbf{e}_i. \quad (4)$$

This expression can explain the origin of ferroelectricity in, e.g., $Ba_2CoGe_2O_7$ [93].

Another example is the cubic perovskite $LaMn_3Cr_4O_{12}$ [94], which looks highly symmetric in its structure and, thus, should not show electric polarization. Both Mn and Cr are magnetic ions, but the magnetic order is a collinear G-type for these two sub-lattices. Then neither the Dzyaloshinskii-Moriya interaction nor the metal-ligand *p-d* hybridization can explain the origin of its tiny polarization. It was demonstrated that in this case the anisotropic symmetric exchange contribution is responsible for the unusual ferroelectric polarization in $LaMn_3Cr_4O_{12}$ [89].

In addition, many phenomenological expressions for both type-I and type-II multiferroics have been proposed in recent years. For example, for the type-II multiferroics $CaMn_7O_{12}$ and $Cu_3Nb_2O_8$, the triple term coupling between polarization component, crystalline axial vector, and magnetic chirality was discussed [95, 96]. And even for some type-I multiferroics, such as strained $BiFeO_3$ (with proper ferroelectricity), $Ca_3MnO_7$ (with improper ferroelectricity), strained $CaMnO_3$, and perovskite superlattices, trilinear or even pentalinear magnetoelectric couplings have been proposed [97-99]. These expressions usually are associated with concrete structural distortion modes, as the collective rotation of oxygen octahedra [100, 101], which determines the sign of the spin-orbit coupling effect (Dzyaloshinskii-Moriya interaction or others). Therefore, the underlying microscopic "glue" remains the spin-orbit-lattice coupling, despite the several complicated manifestations.

*E. Other contributions in magnetoelectric interfaces*

Besides the field effect, there are other routes to obtain magnetoelectricity in heterostructures. The strain effect can play as the mediator between piezoelectricity and magnetostriction. Usually the spin-lattice coupling is responsible for the magnetostriction effect [5, 102]. However, in many cases the spin-orbit coupling is also essential. Although both the spin-orbit and spin-lattice couplings have already been introduced in previous subsections, it is necessary to highlight the complicated physics involved in the process. To obtain a large magnetoelectric response, the magnetocrystalline anisotropy is often utilized, whose microscopic origin is also the spin-orbit coupling. As discussed before, the spin-orbit coupling seriously depends on the crystalline symmetry. Then, in some fine-tuned systems, the strain effect created by the piezoelectric layer may change the crystalline symmetry of the ferromagnetic layer and, thus, change the direction of the magnetocrystalline anisotropy. Then, a 90º switching of the magnetization driven by electric voltage can be obtained straightforwardly [103, 104]. The challenge of this mechanism is to obtain a 180º switching of the magnetization, since the simple magnetocrystalline anisotropic term cannot distinguish between -**S** and +**S**. In spite of these challenges, with a small magnetic field as bias, or via a specially-designed two-step dynamical process, a 180º switching of magnetization can also be obtained in these piezoelectric-ferromagnetic heterostructures with strain-mediated magnetocrystalline anisotropy [105, 106].

Another exotic magnetoelectric phenomenon is the magnetism-controlled charge transfer in the tri-layer superlattice $NdMnO_3/SrMnO_3/LaMnO_3$ [107]. These three manganites are antiferromagnetic but nonpolar. The charge transfers from $NdMnO_3$ and $LaMnO_3$ to $SrMnO_3$ lead to ferromagnetism and, thus, a net magnetization. The key asymmetric charge transfer of these two interfaces creates a net polarization which can be (partially) switched by an external voltage. Furthermore, this polarization can be significantly affected by the magnetic transition as well as by external magnetic fields, since the charge transfer involved depends on the electronic structure which is magnetic-dependent for manganites.

Besides the traditional solid ferroelectrics, ionic liquids can also provide large field effects [108]. Moreover, recent studies have revealed new mechanisms beyond the field effect, which can also tune the magnetism via electric methods. For example, the electric-chemical reactions, i.e. ionic injection/depletion of light ions, can significantly tune the physical properties of materials, including their magnetism [109, 110].

**Perspectives**

Despite the considerable theoretical success in understanding the many magnetoelectric mechanisms acting in multiferroics, there remain several challenges and questions to be solved. For example, the mechanisms based on spin-orbit coupling are typically weak, while those based on spin-lattice coupling are typically insensitive to magnetic fields. Weakness also exists for charge-ordered multiferroics, because of too small band gaps which lead to serious leakage. In recent years, some interesting directions in the field of multiferroics have been explored which may open a new era of magnetoelectricity. Here we list some of these new systems and their novel magnetoelectric physics.

The spin-orbit coupling, as highlighted in previous sections, is certainly one key "glue" to mediate polarization and magnetism. However, the spin-orbit coupling, which is proportional to the atomic number, is weak for $3d$ transition metals as well as for oxygen. To strength the spin-orbit coupling, heavy elements, such as $4d/5d$ transition metals or $4f$ rare earth metals, are possible candidates. However, the multiferroics with these elements have been rarely explored. A possible reason is simply historical: $3d$ transition metal oxides have been far more intensively studied in the past decades, following the development of high-$T_C$ superconducting cuprates and colossal magnetoresistive manganites. Currently, the available $3d$ transition metal components are far more plentiful than the corresponding $4d/5d/4f$ ones, and the physical understanding is also much deeper for $3d$ compounds. However, clearly it is a promising direction to explore new multiferroics in the area of 4$d$/5$d$/4$f$ metals. For example, a recent theoretical prediction proposed that the 3$d$-5$d$ double perovskite $Zn_2FeOsO_6$ could be a room temperature multiferroic with strong ferroelectricity and strong ferrimagnetism [111]. Interestingly, there is a strong magnetoelectric coupling in $Zn_2FeOsO_6$ due to the enhanced spin-orbit coupling effect of the 5$d$ Os element.

For homogeneous systems with both uniform polarization **P** and magnetization **M**, the most common form of the magnetoelectric coupling is **P**$^2$**M**$^2$, which suggests that the reversal of polarization could not lead to a change of magnetization. Recently, it was discovered that there is a novel magnetoelectric coupling of the **PM**$^2$ form, when the parent phase is non-centrosymmetric and non-polar [112]. This magnetoelectric coupling suggests that the reversal of polarization may lead to a flop of the magnetization (a 90° rotation of magnetization). This not only explains the magnetoelectric behavior in the first known multiferroics (i.e., the Ni-$X$ boracite family), but also provides a novel avenue to design/search for new high-performance multiferroics. Similarly, a new form of magnetoelectric coupling was also proposed for the spin-charge coupling in particularly-designed heterostructures: $(\nabla\cdot\mathbf{P})(\mathbf{M}\cdot\mathbf{L})$ where **L** is the antiferromagnetic order, which was expected to achieve the function of magnetization flipping by electric-field [113].

Another interesting direction are the low dimensional multiferroics. Since the discovery of graphene, the zoo of two-dimensional materials has bloomed as a big branch of condensed matter. In early years, most attention on these two-dimensional materials were focused on the semiconducting and optoelectric properties. Only in recent years, more and more intrinsic functions have been re-discovered in two-dimensional materials, including superconductivity, ferroelectricity, and magnetism. Therefore, it is natural to expect the existence of two-dimensional multiferroics, which may provide more convenience for nanoscale magnetoelectric devices. The experience and knowledge learned from three-dimensional magnetoelectric crystals, as reviewed before, can be helpful to search/design low-dimensional multiferroics. For example, the generation of ferroelectricity by noncollinear spin texture has been predicted in Mxene monolayer [114] and the charge-orbital ordering concept has also been implemented in transition-metal halide monolayer to pursue the ferromagnetic ferroelectricity [115]. Furthermore, the concept of a two-dimensional hyper-ferroelectric metal was proposed [116]. In such metallic system, there is an out-of-plane electric polarization which may be switched by an out-of-plane electric field. Since the metallicity is compatible with the strong ferromagnetism, two-dimensional hyper-ferroelectric metals pave a new way to search for the long-sought high temperature ferromagnetic-ferroelectric multiferroics. More two-dimensional multiferroics have been predicted in recent years [117-119]. In summary, low dimensional materials can also host multiferroicity as in the canonical three-dimensional crystals and may display novel physics beyond the three-dimensional counterparts. More efforts, especially from the experimental side, are needed in the future along this direction to verify and manipulate the magnetoelectricity in low dimensions.

## Final Remark

In the half century history of magnetoelectricity and multiferroics, experiments and theories synchronously developed and mutually learned from each other and boosted our knowledge in the field. The discovery of new materials and the revelation of new physics have been greatly accelerated in the 21th century. Benefiting from the enormous efforts accumulated in the past decade, the theories of magnetoelectricity in multiferroics have established a systematic framework involving several key factors within quantum physics and condensed matter physics. Then, the proposed theories of magnetoelectricity are not only addressing the field of multiferroics, but are also widely applicable to the broader field of correlated electronic systems [120, 121]. In this sense, the development of magnetoelectric theories is one of the core physical topics of focus within Condensed Matter Physics in recent times. Certainly additional efforts are much needed to further push forward the physical understanding of this subject and be closer to real applications of these fascinating multiferroic materials.

## Funding

This work was supported by the National Natural Science Foundation of China (11834002, 11674055, and 11825403), the Special Funds for Major State Basic Research (2015CB921700), and the Qing Nian Ba Jian Program. E.D. was supported by the U.S. Department of Energy (DOE), Office of Science, Basic Energy Sciences (BES), Materials Science and Engineering Division.